\documentclass{aa}  
\usepackage{graphicx}
\usepackage{txfonts}
\newcommand{\etal}{et\,al.\ }
\begin{document}
   \title{SDSS\,J212531.92$-$010745.9 - the first definite PG1159 close binary system}
   \author{T. Nagel\inst{1} \and S. Schuh\inst{2} \and D.-J. Kusterer\inst{1} \and
           T. Stahn\inst{2} \and S.D. H\"ugelmeyer\inst{2} \and
           S. Dreizler\inst{2} \and B.T. G\"ansicke\inst{3} \and M.R. Schreiber\inst{4} }
   \offprints{T. Nagel}
   \mail{nagel@astro.uni-tuebingen.de}
   \institute{Institut f\"ur Astronomie und Astrophysik, Eberhard-Karls-Universit\"at T\"ubingen,
              Sand 1, 72076 T\"ubingen, Germany
         \and
             Institut f\"ur Astrophysik, Georg-August-Universit\"at
              G\"ottingen, Friedrich-Hund-Platz 1, 37077 G\"ottingen, Germany 
         \and
             Department of Physics, University of Warwick, Coventry, CV4
              7AL, Great Britain
         \and
             Departamento de Fisica y Meteorologia, Facultad de
             Ciencias, Universidad de Valparaiso, Valparaiso, Chile
             }
   \date{Received xx.xx.xx / Accepted xx.xx.xx}

 
  \abstract
   {}
   {The archival spectrum of SDSS\,J212531.92$-$010745.9 shows not
   only the typical signature of a PG\,1159 star, but also indicates the
   presence of a companion. Our aim was the proof of the binary nature of
   this object and the determination of its orbital period.}
   {We performed time-series photometry of SDSS\,J212531.92$-$010745.9. We
   observed the object during 10 nights, spread over one month, with the 
   T\"ubingen 80\,cm and the G\"ottingen 50\,cm telescopes. We fitted the
   observed light curve with a sine and simulated the light curve of this
   system with the \texttt{nightfall} program. Furthermore, we compared the
   spectrum of SDSS\,J212531.92$-$010745.9 with NLTE models, the results of which 
   also constrain the light curve solution. }
   {An orbital period of 6.95616(33)\,h with an amplitude of 0.354(3)\,mag is derived
   from our observations. A pulsation period could not be detected. For the
   PG\,1159 star we found, as preliminary results from comparison with our NLTE
   models, $T_{\rm eff}$\,$\sim$\,90\,000\,K, $\log g$\,$\sim$\,7.60, and the abundance ratio
   C/He\,$\sim$\,0.05 by number fraction. For the companion we obtained with a mean radius of
   $0.4\pm 0.1\,\rm R_\odot$, a mass of $0.4\pm 0.1\,\rm M_\odot$, and a
   temperature of 8\,200\,K on the irradiated side, good agreement between
   the observed  light curve and the \texttt{nightfall} simulation, but we
   do not regard those values as final. }     
   {}

   \titlerunning{The first definite PG1159 close binary system}
   \authorrunning{T. Nagel at al.}
   \maketitle
%

\section{Introduction}
PG\,1159 stars are hot hydrogen-deficient (pre-)white dwarfs with effective
temperatures between 75\,000 and 200\,000\,K, and $\log g$\,=\,5.5--8.0 (Werner 2001). 
They are in the transition between the asymptotic giant branch (AGB) and cooling white dwarfs.
Spectra of PG\,1159 stars are dominated by absorption lines of He\,{\sc ii}, C\,{\sc iv}
and O\,{\sc vi}. 

Current theory suggests (e.g. Werner 2001) that they are the outcome of a late helium-shell
flash, a phenomenon that drives the currently observed fast evolutionary
rates of three well-known objects (FG~Sge, Sakurai's object, V605
Aql). Flash-induced envelope mixing produces a H-deficient stellar
surface. The photospheric composition then essentially reflects that of the
region between the H- and He-burning shells in the precursor AGB star. The
He-shell flash forces the star back onto the AGB. The subsequent, second
post-AGB evolution explains the existence of Wolf-Rayet central stars of
planetary nebulae and their successors, the PG\,1159 stars. 

\begin{figure}
  \centering
  \includegraphics[width=8cm]{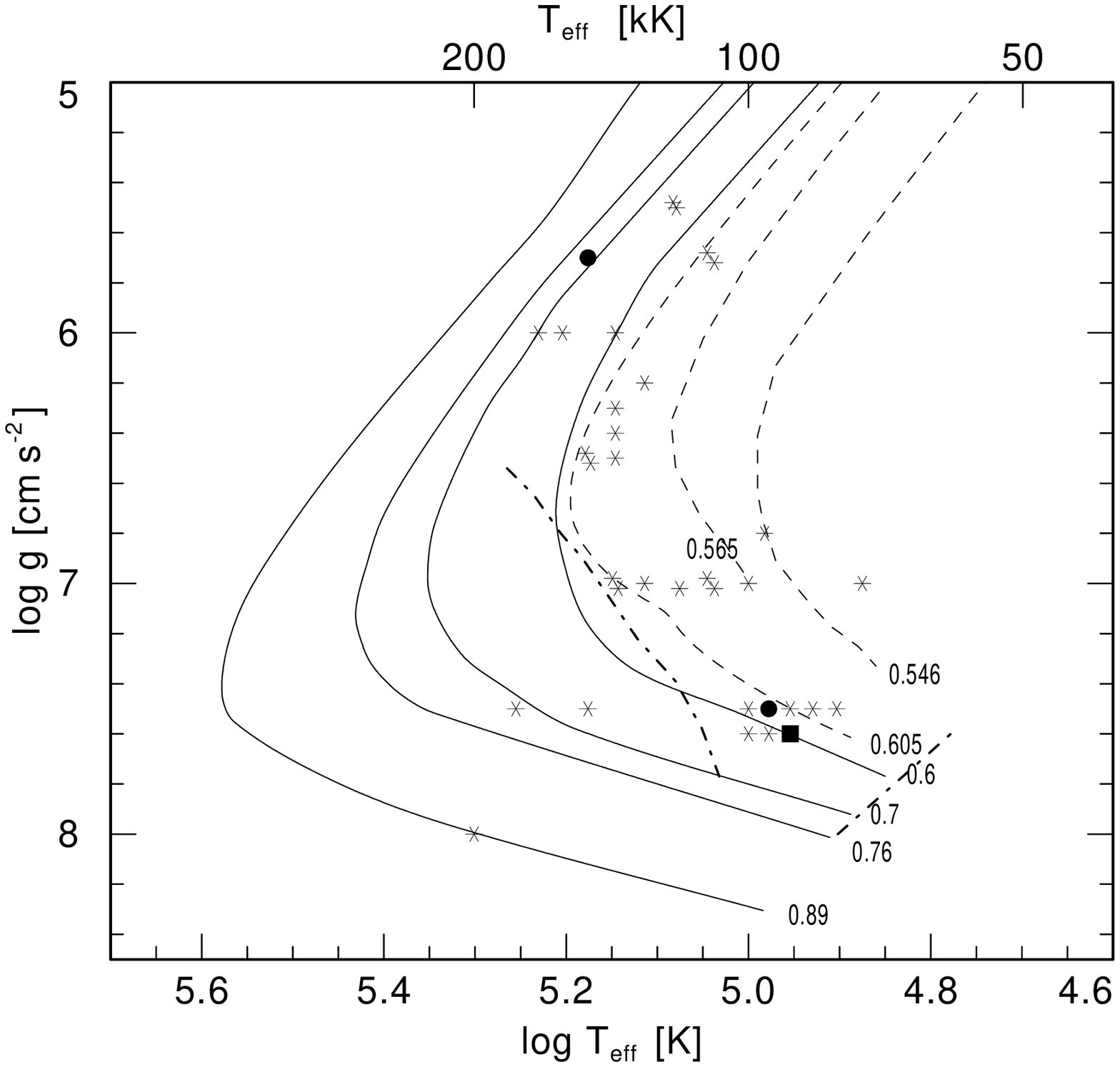}
  \caption{Positions of the known PG\,1159 stars in the log $T_{\rm
  eff}$-$\log g$-diagram. The two known binary systems are shown as black dots,
  the new one is shown as square. Post-AGB evolutionary tracks are taken
  from Sch\"onberner (1983, dashed lines, 0.546\,M$_\odot$ and
  0.565\,M$_\odot$), Bl\"ocker (1995, dashed line, 0.605\,M$_\odot$), and
  Wood \& Faulkner (1986, solid lines) (labels: mass in M$_\odot$). The
  dashed-dotted lines represent the theoretical red (Quirion \etal 2004)
  and blue edge (Gautschy priv.comm.) of the GW  Vir instability strip.}  
  \label{pg1159stars}
\end{figure}

Currently, 37 PG\,1159 stars are known. Figure\,\ref{pg1159stars} shows their position in a
log\,$T_{\rm eff}$-$\log g$-diagram. Two of them have been found to be
binary stars. These are NGC\,246 (e.g. Bond \& Ciardullo 1999), which is a
resolved visual binary, and PG\,2131+066 (Wesemael \etal 1985). Concerning
the latter, it is still unclear whether it is a close binary (Paunzen \etal
1998) or a resolved visual binary with an M2V star as companion (Reed \etal 2000). 

\section{The spectrum of SDSS\,J212531.92$-$010745.9}
\begin{figure*}
  \centering
  \includegraphics[width=17cm,height=9cm]{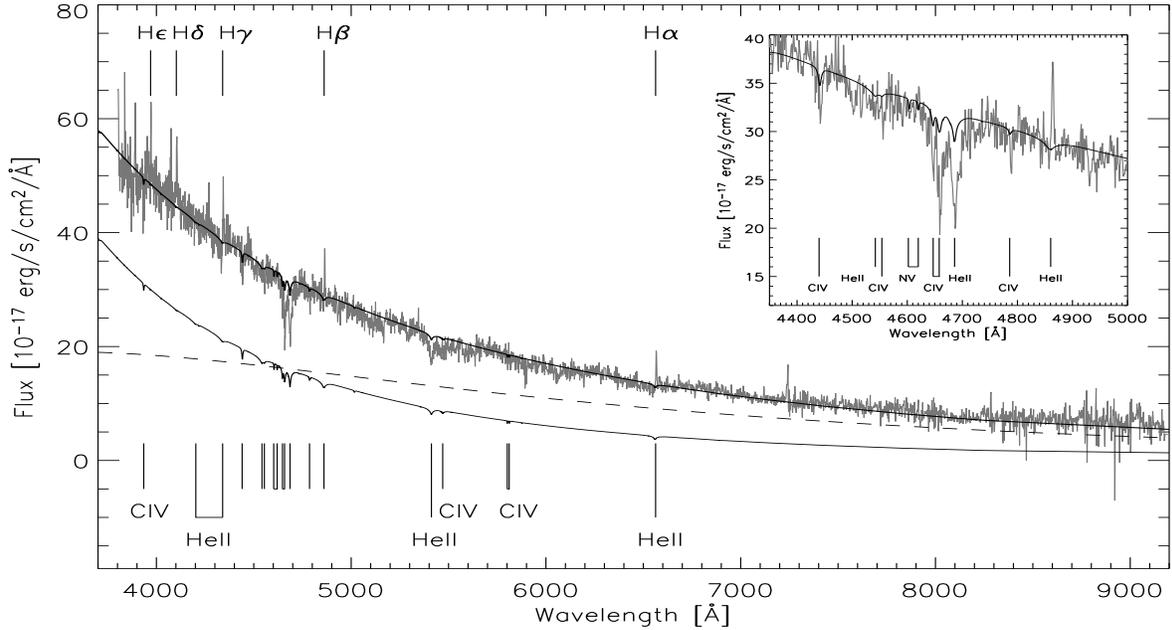}
  \caption{Spectrum of SDSS\,J212531.92$-$010745.9 (gray line, exposure time 3703\,s), a PG\,1159 NLTE
  model spectrum with $T_{\rm eff}$\,=\,90\,000\,K, $\log g$\,=\,7.60, and
  C/He\,=\,0.05, N/He\,=\,0.01 (thin black line), a blackbody spectrum 
  with $T$\,=\,8\,200\,K (dashed line), representing the
  contribution from the irradiated companion, and the total model spectrum (thick
  black line). The Balmer series in emission (top), belonging to
  the companion, and some helium and carbon lines (bottom), belonging
  to the PG\,1159 star, are marked.}  
  \label{spectrum}
\end{figure*}

The spectrum of SDSS\,J212531.92$-$010745.9 (u=17.15, g=17.54, r=17.75,
i=17.79, z=17.83), taken on Sept. 6th 2002, is from the Sloan Digital
Sky Survey (SDSS) archive Data Release (DR) 4. The spectrum shows significant features
that are typical for PG\,1159 stars, for example the strong C\,{\sc iv}
absorption lines at $4650-4700\AA$ and He\,{\sc ii} at $4686\AA$ (Fig.\,\ref{spectrum}).  
Furthermore, the spectrum shows features which indicate the presence of a
companion. The Balmer series of hydrogen is seen in emission, H$_\alpha$ -
H$_\delta$ can clearly be identified. This is probably due to a cool
companion which is heated up by irradiation from the hydrogen-deficient PG\,1159 star. 

Figure\,\ref{spectrum} shows the observed spectrum ($t_{\rm exp}=3703\,\rm s$) of
SDSS\,J212531.92$-$010745.9. Overlayed are a PG\,1159 NLTE model spectrum
with $T_{\rm eff}$\,=\,90\,000\,K, $\log g$\,=\,7.60, C/He\,=\,0.05, and N/He\,=\,0.01, a
blackbody model spectrum with $T$\,=\,8\,200\,K for the irradiated
companion, and the sum of the two model spectra. The parameters of both
stellar components are estimates obtained from a qualitative comparison of
our NLTE models to the single SDSS spectrum. Detailed parameters for both stars  
need to be derived from a full two-component analysis of orbital 
phase resolved spectroscopy. The effective temperature in particular may be
lower or higher by 20\,000\,K. The surface temperature of the companion's irradiated
side was also constrained with \texttt{nightfall} simulations, see below. 

The overall shape of the observed spectrum is well fitted with the
combination of a PG\,1159 star and a cool, irradiated companion, but
especially the C\,{\sc iv} spectral lines of the PG\,1159 model atmosphere
are not strong enough. There is another PG\,1159 star showing this
phenomenon (H\"ugelmeyer \etal, in prep.), and also none of the deep
absorption lines which some DO white dwarfs show can be fitted (e.g. Werner \etal 1995).

The spectral signatures of an A star, as one would expect for the companion with
8\,200\,K surface temperature at the irradiated side, cannot be seen in
the observation. This may be because the irradiation from the PG\,1159
leads to a temperature inversion in the upper layers of the companion's
atmosphere up to $\tau_{\rm Ross}=1$, which causes the observed emission line spectrum
(Barman \etal 2004). 

\section{Photometry of SDSS\,J212531.92$-$010745.9}
\begin{figure*}
  \centering
  \includegraphics[width=17cm,height=8cm]{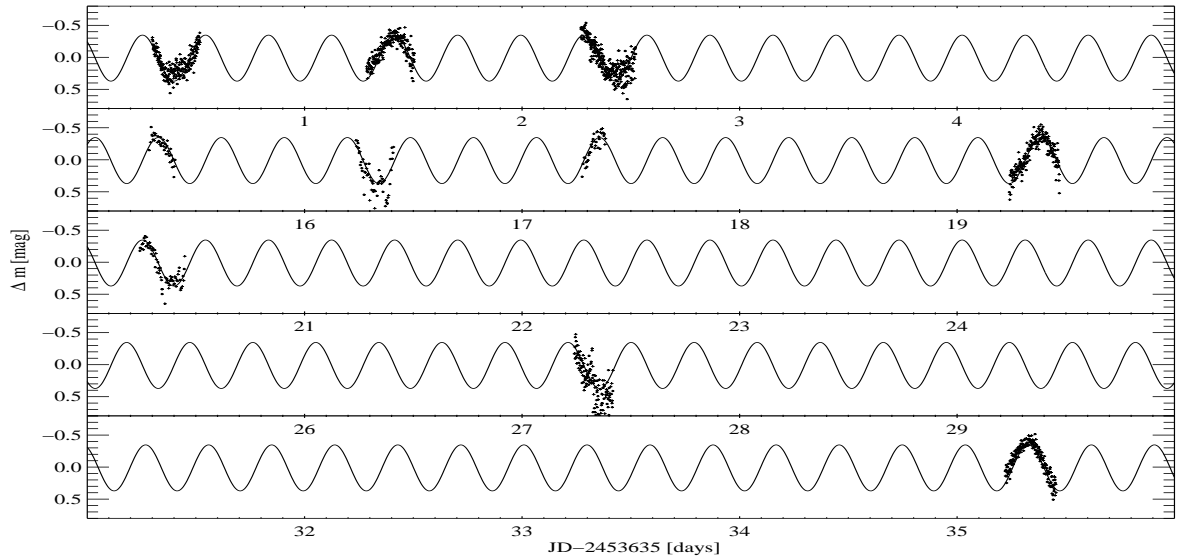}
  \caption{Light curve of all nights, overplotted the best sine fit with
    a period of 6.95616\,h. }
  \label{lightcurve}
\end{figure*}
\begin{table}
\centering
\caption{Observation log. All observations are performed with clear filter.}
\begin{tabular}{rrrcc}
  \hline
  \hline
  \noalign{\smallskip}
  Date  &  t$_{\rm exp}[s]$&  t$_{\rm cycle}[s]$  &  Duration$[s]$  &  Telescope\\
  \noalign{\smallskip}
  \hline
  \noalign{\smallskip}
  2005/09/21 &  90 & 98 &  18900 & 80\,cm\\
  2005/09/22 &  90 & 98 &  18899 & 80\,cm\\
  2005/09/23 &  90 & 98 &  21758 & 80\,cm\\
  2005/09/23 & 180 & 194&  14873 & 50\,cm\\
  2005/10/06 & 240 & 248&  10202 & 50\,cm\\
  2005/10/07 & 240 & 246&  14897 & 50\,cm\\
  2005/10/08 & 240 & 248&   9298 & 50\,cm\\
  2005/10/10 &  90 & 98 &  19852 & 80\,cm\\
  2005/10/11 & 240 & 248&  17872 & 50\,cm\\
  2005/10/18 &  90 & 98 &  16532 & 80\,cm\\
  2005/10/26 &  90 & 98 &  20095 & 80\,cm\\
  \noalign{\smallskip}
  \hline
\end{tabular}\label{tab_obs}
\end{table}

Photometric observations of SDSS\,J212531.92$-$010745.9  were
performed during 10 nights (Tab.\,\ref{tab_obs}) using the T\"ubingen
80\,cm f/8 telescope with an \mbox{SBIG ST-7E} CCD camera and the G\"ottingen
50\,cm f/10 telescope with an \mbox{SBIG STL-6303E} CCD camera. To achieve
good time resolution we chose clear filter exposures with a binning of 2x2
pixels to reduce readout time. The exposure time was t$_{\rm exp}$=90\,s
for the observations with the 80\,cm telescope. In the case of the 50\,cm
telescope, the exposure time was t$_{\rm exp}$=180\,s and t$_{\rm
  exp}$=240\,s. The observing conditions were good during the nights,
considering that the telescopes are located in the cities of T\"ubingen and
G\"ottingen.   

All images were bias and dark current corrected, then aperture 
photometry was performed using our IDL software TRIPP (Time Resolved
Imaging Photometry Package, Schuh \etal 2003). The relative flux of the
object was calculated with respect to the same two comparison stars (SDSS
J212530.60-010921.0 and SDSS J212528.83-010828.5) for all nights, which
were tested for stability. The resulting light curve is displayed in
Fig.\,\ref{lightcurve}.  

To analyse the combined light curve of all nights, we used CAFE (Common
Astronomical Fit Environment, G\"ohler, priv. comm.), a collection of
routines written in IDL. The brightness variation is probably caused by a
reflection effect. The companion is, due to the small separation, heated up
on one side by irradiation from the PG\,1159 star, and the orbital motion
then leads to a variable light curve. 
We fitted the combined light curve of all nights with a sine, achieving
best results for a period of 6.95616(33)\,h (Fig.\,\ref{lightcurve}). The
observed variability has a mean amplitude of 0.354(3)\,mag. 

\begin{figure}
  \centering
  \includegraphics[width=9cm]{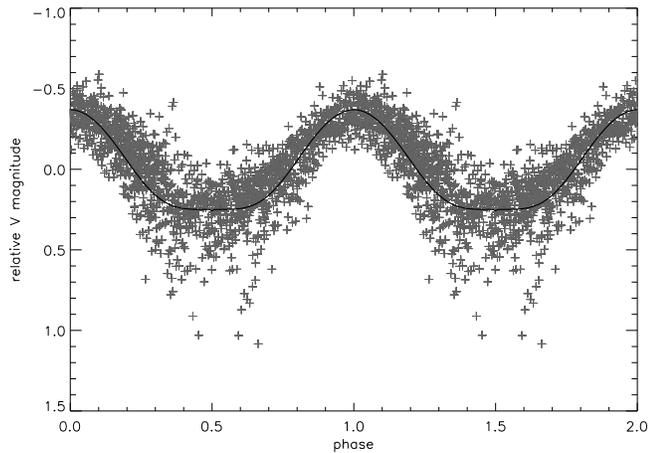}
  \caption{Simulated light curve of a binary system, consisting of a PG\,1159
  star ($T_{\rm eff}$\,=\,90\,000\,K) and an M dwarf ($T_{\rm
  eff}$\,=\,3\,500\,K, heated up to 8\,200\,K), calculated with \texttt{nightfall} (black line) 
  and the observed light curve of all nights, folded onto the orbital
  period. The shape of the light curve is well resampled, especially the
  broad dips.} 
  \label{nightfall}
\end{figure}

To check if the observed light curve can be explained by a PG\,1159 star
and an irradiated companion and for an impression of what the system
geometry might look like we simulated the light curve of the binary system
for an orbital period of 6.95616\,h with the program \texttt{nightfall}. 
Figure\,\ref{nightfall} shows the simulated and observed light curves of
all nights, folded onto the orbital period. For the PG\,1159 star we assumed
$T_{\rm eff}$\,=\,90\,000\,K, a mass of 0.6\,$\rm M_\odot$ and a radius of
0.1\,$\rm R_\odot$.  
For the companion we varied the mass from 0.1\,$\rm M_\odot$ to 0.7\,$\rm
M_\odot$. We found that the observed light curve can be reproduced best
with an M dwarf with an effective temperature of $3\,500\pm 150\,\rm K$, 
a mean radius of $0.4\pm 0.1\,\rm R_\odot$ and a mass of about $0.4\pm
0.1\,\rm M_\odot$. For the inclination of this system we obtained 
$70\pm 5\,^\circ$.  
Due to the irradiation by the PG\,1159 star the surface of the companion
would be heated up to a surface temperature of 8\,200\,K, which, in
combination with the PG\,1159 star, reproduces the overall shape of the
observed spectrum quite well, as can be seen in Fig.\,\ref{spectrum}. The
broad dip at the minimum of the light curve is well reproduced by this
system configuration, too. 
In Table \ref{paras} we list all stellar and system parameters assumed and
derived.  
We found that ellipsoidal variation due to geometrical deformation of
the stars cannot generate the observed light curve. In the above
configuration, calculated by nightfall according to the geometry in
Djurasevic 1992, the equatorial radius of the M dwarf is only 4.5\,\%
larger than its polar radius, and the PG\,1159 star is not affected by
deformation above the numerical limit of \texttt{nightfall}.

Because the object is positioned in the GW\,Vir instability strip
(Fig.\,\ref{pg1159stars}), we also looked for pulsation periods below two
hours in the light curve of SDSS\,J212531.92$-$010745.9. Therefore, we
calculated a Lomb-Scargle periodogram (Scargle 1982) for the night with the
best S/N (2005/10/26). But observational noise precludes detection of any
periodicity with amplitudes below about 50\,mmag. The amplitudes of
pulsating PG\,1159 stars normally are of the order of a few percent of a
magnitude, and SDSS\,J212531.92$-$010745.9 is fainter than HE1429-1209, for
which we recently discovered pulsation with the T\"ubingen 80\,cm telescope
(Nagel \& Werner 2004). Our 80\,cm telescope might therefore just be too
small to detect pulsation below 50\,mmag in this case.  

\begin{table}
\centering
\caption{Stellar and system parameters of SDSS\,J212531.92$-$010745.9,
  assumed (normal font) or derived from comparison with NLTE model spectra (boldface),
  photometric analysis (*) and nightfall simulation (italic).}  
\begin{tabular}{rrrr}
  \hline
  \hline
  \noalign{\smallskip}
  Parameter                      & PG\,1159            & Companion         & System \\
  \noalign{\smallskip}
  \hline
  \noalign{\smallskip}
 $T_{\rm eff}\,\,[\rm K]$        & {\bf $\sim$90\,000} & 3\,500$\pm$150    &  \\  
 $T_{\rm eff,irr} \,\, [\rm K]$  &                     & {\it 8\,200}      &  \\
 $\log g  \,\,[\rm cm/s^2]$      & {\bf $\sim$7.6}     &                   &  \\
 $m  \,\,[\rm M_\odot]$          & 0.6                 & {\it 0.4$\pm$0.1} & 1.0$\pm$0.1 \\
 $r  \,\,[\rm R_\odot]$          & 0.1                 & {\it 0.4$\pm$0.1} &  \\
 $P_{\rm orb}  \,\,[\rm h]$      &                     &                   & 6.95616(33) *\\
 $\Delta m  \,\,[\rm mag]$       &                     &                   & 0.354(3)  * \\
 $a  \,\,[\rm R_\odot]$          &                     &                   & 1.85\\
 $i  \,\,[^\circ]$               &                     &                   & {\it 70$\pm$5}\\
 \noalign{\smallskip}
  \hline
\end{tabular}\label{paras}
\end{table}

\section{Conclusions}
\begin{enumerate}
\item The spectrum of SDSS\,J212531.92$-$010745.9 from DR4 of the Sloan Digital
  Sky Survey shows the signature of a PG\,1159 star plus emission from a
  cool irradiated companion. 
\item We performed time-series photometry during 10 nights with the
  T\"ubingen 80\,cm and the G\"ottingen 50\,cm telescopes and detected a
  period of 6.95616(33)\,h with an amplitude of 0.354(3)\,mag. This
  represents the orbital period of the binary system. Thus,
  SDSS\,J212531.92$-$010745.9 is the first close PG\,1159 binary without any
  doubts. 
\item From a first comparison with NLTE model spectra we derived, as preliminary
  results, an effective temperature of 90\,000\,K, $\log g\,\sim$\,7.60 and
  the abundance ratio C/He\,$\sim$\,0.05 for the PG\,1159 component.   
  A detailed, quantitative NLTE spectral analysis of the PG\,1159 star
  and the irradiated companion has to be done next. We will report on the
  results in a subsequent paper.
\item We simulated the light curve of the binary system with an orbital
  period of 6.95616\,h using \texttt{nightfall}. A good agreement with the
  observed light curve was obtained for a mean radius of $0.4\pm 0.1\,\rm
  R_\odot$, a mass of \,$0.4\pm 0.1\,\rm M_\odot$ and a temperature of the
  irradiated surface of about 8\,200\,K for the companion.   
\end{enumerate}
To determine the system parameters more precisely, high-resolution 
phase-resolved spectroscopy of SDSS\,J212531.92$-$010745.9 is necessary.
It should then be possible to derive both the companion's variable 
light contribution to the overall spectrum as well as dynamical 
masses for both components from radial velocity measurements of their
distinct line systems.

\begin{acknowledgements}
      We thank T.-O. Husser, R. Lutz and E. Nagel for supporting the
      observations. We acknowledge the use of {\texttt CAFE 5.1}, an
      astronomical fit enviroment, written by Eckart G\"ohler. We
      acknowledge the use of the \texttt{nightfall} program for the 
      light curve synthesis of eclipsing binaries written by Rainer Wichmann
      (http://www.lsw.uni-heidelberg.de/$\sim$rwichman/Nightfall.html). BTG
      was supported by a PPARC Advanced Fellowship. 
\end{acknowledgements}

\end{document}